\documentclass[11pt,letterpaper]{article}
\usepackage[margin=1in]{geometry}
\usepackage[T1]{fontenc}
\usepackage{latexsym,graphicx}
\usepackage{amsmath,amssymb,amsfonts,amsthm,mathtools}
\usepackage{array,booktabs,enumitem,tikz,algorithm,algorithmicx}
\usepackage[noend]{algpseudocode}
\usetikzlibrary{arrows.meta,positioning,calc,fit,backgrounds}
\usepackage{caption}
\usepackage{setspace,titlesec,tcolorbox}
\usepackage[backend=biber,style=alphabetic,natbib=true]{biblatex}
\usepackage[colorlinks,linkcolor=blue]{hyperref}
\usepackage[nameinlink,noabbrev]{cleveref}
\usepackage{csquotes}
\usepackage{mathtools}
\makeatletter
\providecommand{\theHALG@line}{\thealgorithm.\arabic{ALG@line}}
\makeatother

\Crefname{claim}{Claim}{Claims}
\Crefformat{equation}{#2(#1)#3}

\numberwithin{figure}{section}
\newtheorem{theorem}{Theorem}
\newtheorem{lemma}{Lemma}
\newtheorem{claim}{Claim}

\newtheorem{proposition}{Proposition}

\theoremstyle{definition}

\renewcommand{\textsc}[1]{{\textup{\scshape#1}}}
\titleformat{\paragraph}[runin]
  {\normalfont\normalsize\bfseries}{}{0pt}{}
\titlespacing*{\paragraph}{0pt}{1.5ex}{1em}

\definecolor{mygreen}{RGB}{10,110,230}
\definecolor{myred}{RGB}{10,110,230}
\hypersetup{
 colorlinks=true,
 citecolor=mygreen,
 linkcolor=myred,
 urlcolor=black,
 pdftitle={Generalized Sorting via Predecessor Propagation},
 pdfsubject={A detailed research draft of the four-thirds query bound}
}

\DeclarePairedDelimiter{\paren}{\lparen}{\rparen}

\DeclarePairedDelimiter{\set}{\lbrace}{\rbrace}
\DeclarePairedDelimiter{\card}{\lvert}{\rvert}

\newcommand{\tO}{\widetilde O}

\newcommand{\FourCycleSort}{\mathsf{FourCycleSort}}

\newcommand{\Reduce}{\mathsf{Reduce}}
\newcommand{\round}{\mathsf{round}}

\newcommand{\PredictionSort}{\mathsf{PredictionSort}}
\newcommand{\Epred}{\widetilde{E}}
\newcommand{\ProcessBucket}{\mathsf{ProcessBucket}}
\newcommand{\Peel}{\mathsf{Peel}}
\newcommand{\ExpandWitness}{\mathsf{ExpandWitness}}
\newcommand{\LocalPredict}{\mathsf{LocalPredict}}
\newcommand{\Predict}{\mathsf{Predict}}
\newcommand{\Ehkk}{\Epred_{\mathsf{HKK}}}
\newcommand{\err}{\operatorname{err}}
\newcommand{\LComment}[1]{\State \textcolor{black!50}{// #1}}

\definecolor{queryblue}{RGB}{31,102,157}
\definecolor{truthgreen}{RGB}{26,115,77}
\definecolor{errorred}{RGB}{174,52,43}
\definecolor{defergold}{RGB}{157,104,22}
\tikzset{vertex/.style={circle,draw=black!75,fill=white,minimum size=7mm,inner sep=1pt},
edge/.style={draw=black!65,line width=.8pt},
query/.style={-{Stealth[length=2mm]},draw=queryblue,line width=1.1pt},
truth/.style={-{Stealth[length=2mm]},draw=truthgreen,line width=1.1pt,dashed},
wrong/.style={-{Stealth[length=2mm]},draw=errorred,line width=1.2pt},
flow/.style={-{Stealth[length=2mm]},draw=black!65,line width=.7pt},
box/.style={draw=black!55,rounded corners=2pt,fill=black!2,align=center,
             inner sep=6pt,font=\small}}
\allowdisplaybreaks[1]
\title{\textbf{Generalized Sorting via Four-Cycles}}
 \author{Soheil Behnezhad\thanks{Supported in part by NSF CAREER Award CCF 2442812, a Sloan Fellowship, and a Google Research Award.}\\{\em Northeastern University} \and Shane Ferrante\footnotemark[1]\\{\em Northeastern University}}
\date{}

\begin{document}
\maketitle

\thispagestyle{empty}

\begin{abstract}
\setlength{\parskip}{0.2cm}
We study the {\em generalized sorting} problem where only a subset of the pairs can be compared. Equivalently, we are given an $n$-vertex $m$-edge graph $G=(V, E)$ where the vertex set $V$ is to be sorted, and the edge set $E$ indicates possible comparisons. The goal is to sort $V$ with as few (adaptive) queries to $E$ as possible.


When $G$ is the complete graph, the problem is equivalent to standard sorting for which $\Theta(n \log n)$ queries are sufficient and necessary. All classical sorting algorithms crucially rely on the ability to compare any pair and do not imply an $o(n^2)$-query algorithm in the general case. This was improved 15 years ago by Huang, Kannan, and Khanna [FOCS'11] who achieved a bound of $\widetilde{O}(n\sqrt{n})$. Kuszmaul and Narayanan [FOCS'21] later gave a bound of $\widetilde{O}(\sqrt{m n})$, which is more efficient for sparse graphs, but matches the $\widetilde{O}(n\sqrt{n})$ barrier for dense graphs.

In this paper, we present an algorithm for generalized sorting that makes at most $\widetilde{O}(n^{4/3})$ queries, therefore breaking the longstanding $\widetilde{O}(n\sqrt{n})$ barrier for all graphs. The key to our result is to query edges that belong to many (specific types of) four-cycles. We show that querying such edges allows the algorithm to verify the direction of many other edges (specifically, their opposing edges in the four-cycles) and thus achieve a much better amortized query cost.
\end{abstract}


\clearpage
\pagenumbering{arabic}
\section{Introduction}

The standard sorting problem assumes that any pair of elements can be compared. A well-studied generalization, known as the {\em generalized sorting} problem \cite{HKK}, removes this freedom. We are given a simple
undirected graph $G=(V,E)$ on $n$ vertices and $m$ edges. The vertices have an unknown total order $\prec$, and a query on $\{u,v\}\in E$ reveals whether $u\prec v$ or $v\prec u$. Queries outside $E$ are forbidden. We are promised that consecutive vertices in the true order are joined by an edge. Equivalently, directing every allowed edge according to $\prec$ produces an acyclic graph with a directed Hamiltonian path. The task is to recover this order using few queries.

\paragraph{Background:} If $G$ is the complete graph, then the problem is equivalent to classical comparison-based sorting for which $\Theta(n \log n)$ queries are sufficient and necessary. Another well-studied special case of the generalized sorting problem is when $G$ is the complete bipartite graph, which is known as the {\em nuts-and-bolts matching} problem. For this variant, too, a long line of work eventually achieved the optimal query complexity of $\Theta(n \log n)$~\cite{Rawlins,ABFKNO,BF,Bradford,ABF,KMSjournal}.

The situation is much less understood for general graphs. The trivial algorithm is to use up to $O(n^2)$ queries to learn the direction of every edge in the graph to discover the directed Hamiltonian path.  \citet{HKK} were the first to break this bound. They proved 15 years ago that there is a randomized generalized sorting algorithm with query complexity $\widetilde{O}(n\sqrt{n})$. Roughly speaking, their algorithm iteratively discovers ordered sets of length $\Theta(\sqrt{n})$ using $\widetilde{O}(n)$ queries. Repeating this $O(\sqrt{n})$ times achieves the claimed $\widetilde{O}(n\sqrt{n})$ query complexity. A more recent algorithm by \citet{KN}---building on the work of \citet{LRSZ}---achieves a bound of $\widetilde{O}(\sqrt{mn})$. Their algorithm keeps querying the edges of $G$ uniformly at random until these queries become very unlikely to discover ``inversions'', at which point they switch to the prediction-based algorithm of \cite{LRSZ}.

The algorithms of \cite{HKK} and \cite{KN} achieve the same
bound of $\widetilde{O}(n\sqrt{n})$ for general graphs using
very different approaches. This is also a natural bound
for various other graph problems. For example,
$\widetilde{\Theta}(n\sqrt{n})$ is provably the right
bound for $(\Delta+1)$ vertex coloring under neighbor
and pair queries \cite{ACK}, and for one-sided testing
of graph isomorphism under adjacency-matrix queries
\cite{FM08}. The same bound is optimal in quantum query complexity,  for problems such as graph connectivity, minimum spanning
tree \cite{DHHM}, and testing graph planarity
or acyclicity \cite{CK}, among many others. It is therefore natural to ask:

\begin{quote}
    {\em Is $\widetilde{O}(n\sqrt{n})$ the right bound for generalized sorting?}
\end{quote}

\paragraph{Our result:} In this paper, we break through the longstanding $\widetilde{O}(n\sqrt{n})$ barrier for generalized sorting, showing that it is indeed not the right bound. More specifically, we prove that:

\begin{tcolorbox}
\vspace{0.1cm}
\begin{theorem}\label{thm:main}
Generalized sorting on an $n$-vertex 
graph admits a randomized polynomial-time algorithm that succeeds with probability $1-1/n$  and makes at most
$
 O\!\left(
  n^{4/3}\log^{4/3} n\right)
$
queries.
\end{theorem}
\end{tcolorbox}

Our key ingredient to prove \Cref{thm:main} is to query edges that belong to many (specific types of) length-four cycles that are abundant in dense graphs. We show that querying such edges allows the algorithm to verify the direction of many other edges (specifically, their opposing edges in the four-cycles) and thus achieve a much better amortized query cost. We give an overview of this idea and the overall proof in \Cref{sec:techniques}.

\paragraph{Further related work:}
Generalized sorting builds on earlier work on restricted comparisons and
nonuniform comparison costs. \citet{CFGKRS} first introduced the
priced-information framework, evaluating query strategies against the
cost of a cheapest certificate. Within this framework, \citet{GK} studied sorting and selection with structured comparison
costs, and \citet{KK} studied selection with monotone
comparison costs.

Another related line of work studies monotonicity testing, including
whether an array of numbers is sorted or far from sorted, using few
queries. Each query reveals an array entry, and the distance to sortedness
is the minimum fraction of entries that must be changed to make the array
nondecreasing. Sortedness testing, tolerant testing, and extensions to
partially ordered domains have been studied extensively; see,
e.g.,~\cite{EKKRV,FLNRRS,Fischer,ACCL,PRR,SSdistance,RaskhodnikovaSurvey,Belovs,RonSurvey,Yoshida}
and the references therein.
\section{Our Techniques}\label{sec:techniques}

Throughout this overview, we suppress logarithmic factors. We also do not focus on running time, and only attempt to bound the number of queries. All randomized guarantees hold with high probability; the formal statements and proofs appear in the subsequent sections.

\paragraph{Inversions, and an overview of \cite{KN}:} Let us start by reviewing the concept of inversions, introduced in the work of \cite{HKK} and further utilized by \cite{KN} to achieve their $\tO(\sqrt{m n})=\tO(n\sqrt{n})$ query algorithm.

Let $\Pi$ be the set of all permutations of $V$ that do not contradict any of the queries made so far. For any unqueried edge $\{u, v\} \in E$ let its {\em implied direction} be the direction it takes in the majority of $\Pi$ (breaking ties arbitrarily).  We say that an edge $e \in E$ is an {\em inversion} if its true direction differs from its implied direction. Now suppose that we query $r$ edges uniformly at random. If an inversion is found, it invalidates at least half of the permutations in $\Pi$. In this case, we update $\Pi$ and all the implied directions. Since $|\Pi| = n!$ initially, this happens at most $\log_2 (n!) = O(n \log n)$ times, and the total cost of these queries is at most $O(r \cdot n \log n) = \widetilde{O}(nr)$. If no inversion is found, then the total number of inversions must be at most $O(m/r)$ w.h.p., and so the implied directions predict the true directions highly accurately. At this point, \cite{KN} plug in the algorithm of \cite{LRSZ} which recovers the true order from
a prediction with $w$ errors using $O(n\log n+w)$ queries. The
combined query cost is
$
                         \tO\!\left(nr+m/r \right).
$
Choosing $r = \sqrt{m/n}$, gives the
$\tO(\sqrt{mn}) = \tO(n\sqrt{n})$ bound.

\begin{figure}
\centering
\begin{tikzpicture}[font=\small,vertex/.append style={minimum size=5.5mm}]
\node[vertex] (u1) at (0,2.3) {$u$};
\node[vertex] (v1) at (2.3,2.3) {$v$};
\node[vertex] (b1) at (0,0) {$b$};
\node[vertex] (a1) at (2.3,0) {$a$};
\draw[query] (u1)--node[above=5pt] {$e$}(v1);
\draw[query] (u1)--node[rotate=90,anchor=south,inner sep=8pt,font=\scriptsize] {predicted}(b1);
\draw[query] (a1)--node[rotate=90,anchor=north,inner sep=8pt,font=\scriptsize] {predicted}(v1);
\draw[query] (a1)--node[below=5pt,font=\scriptsize] {opposite edge}(b1);

\draw[-{Stealth[length=3mm]},draw=black!70,line width=1pt]
  (3.35,1.15)--node[above=6pt] {after 3 queries}(5.25,1.15);

\begin{scope}[xshift=6.3cm]
\node[vertex] (u2) at (0,2.3) {$u$};
\node[vertex] (v2) at (2.3,2.3) {$v$};
\node[vertex] (b2) at (0,0) {$b$};
\node[vertex] (a2) at (2.3,0) {$a$};
\draw[wrong] (v2)--node[above=5pt,font=\scriptsize] {inversion}(u2);
\draw[query] (u2)--node[rotate=90,anchor=south,inner sep=8pt,font=\scriptsize] {confirmed}(b2);
\draw[query] (a2)--node[rotate=90,anchor=north,inner sep=8pt,font=\scriptsize] {confirmed}(v2);
\draw[truth] (a2)--node[below=5pt,font=\scriptsize,align=center]
  {inferred\\(not queried)}(b2);
\end{scope}
\end{tikzpicture}
\caption{A four-cycle witness.}
\label{fig:witness}
\end{figure}
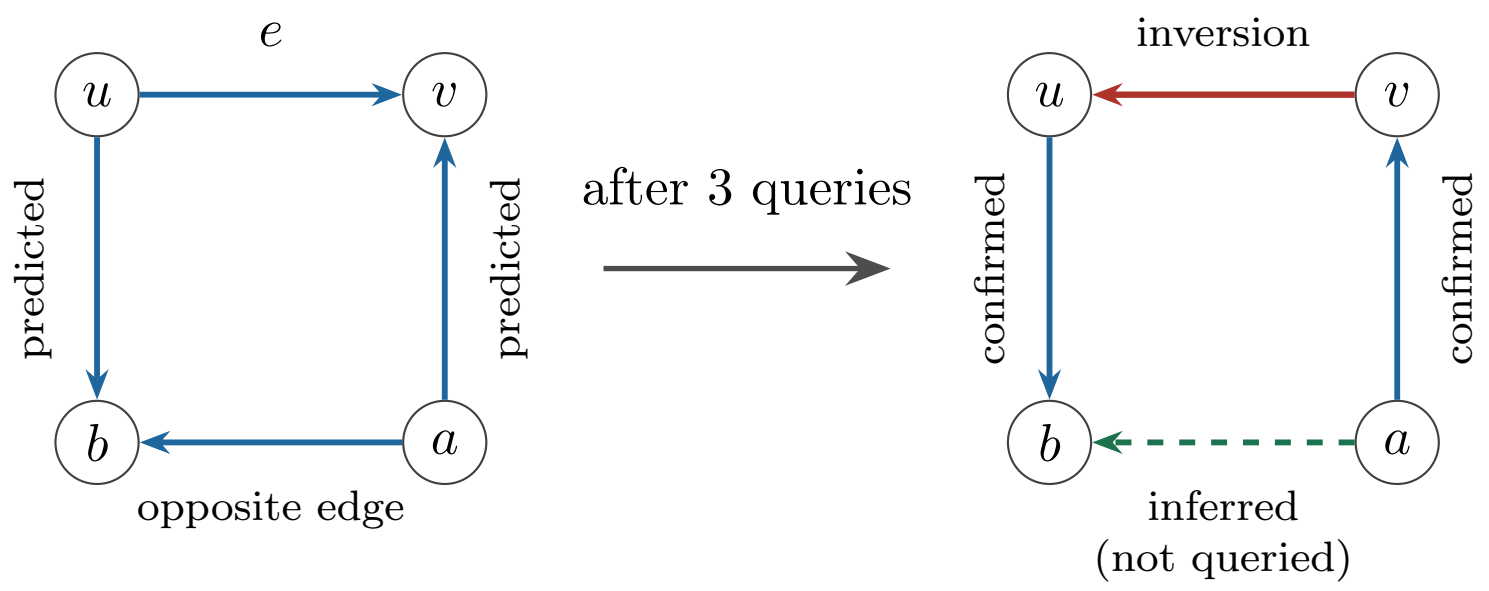

\paragraph{Key ingredient: inversions and witnesses:} In the argument above, inversions are valuable because they reduce the solution space. Our algorithm also makes progress by finding inversions, but does so in a very different way. We call a cycle on 4 vertices $a, b, u, v$ a {\em witness} if its implied edge directions are exactly as in \Cref{fig:witness}. Now suppose that 3 queries confirm the directions of $(u, b)$, $(a, v)$, but reveal $(u, v)$ to be an inversion. In this case, we automatically infer $(a, b)$ to be predicted correctly.

At first, it may not seem that useful to make 3 queries only to confirm the direction of 4 edges. And indeed, that's not what we do. We show that if the graph is dense enough, then an average edge belongs to many witnesses, and so finding an inversion and querying its incident edges verifies the direction of a significant number of opposing edges. The effect of this (despite having a higher cost for querying all incident edges) is much more drastic than reducing the solution space by a constant. Let us describe this intuition further by considering random graphs first.


\paragraph{Existence of many witnesses (random graphs).}
First, observe that we can equivalently view witnesses as four-cycles
in a bipartite graph. To see this, take two copies of
every vertex $v\in V$, denoted by $v^-$ and $v^+$.
For every predicted edge $(u,v)$, we add an undirected
edge between $u^-$ and $v^+$. This gives us a bipartite
graph $H$ with $n$ vertices on each side and $m$ edges.
As illustrated below, every
four-cycle in $H$ corresponds exactly to a witness
in the original graph: a four-cycle on
$u^-,a^-,v^+,b^+$ corresponds to the four predicted
edges $(u,v),(u,b),(a,v),(a,b)$.

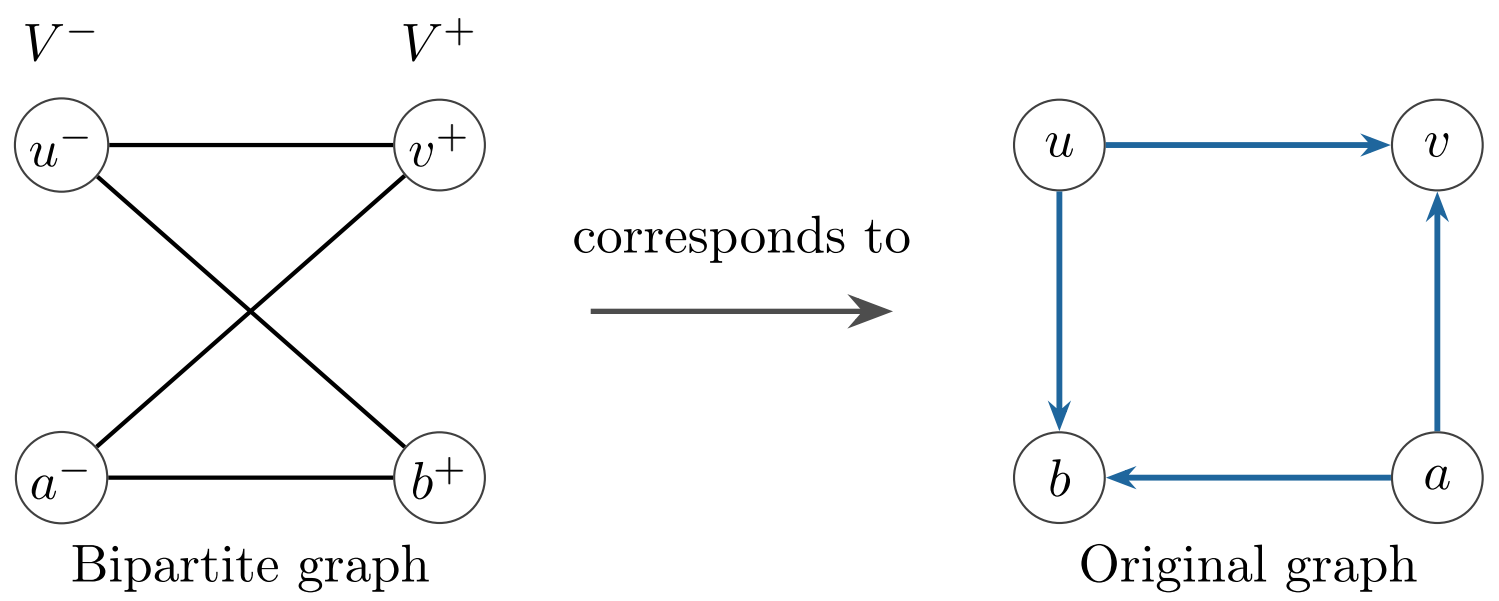
\begin{figure}[h]
\centering
\begin{tikzpicture}[font=\small,
    vertex/.append style={minimum size=6mm}]

\node[vertex] (um) at (0,2.2) {$u^-$};
\node[vertex] (am) at (0,0) {$a^-$};
\node[vertex] (vp) at (2.5,2.2) {$v^+$};
\node[vertex] (bp) at (2.5,0) {$b^+$};

\draw[thick] (um)--(vp);
\draw[thick] (um)--(bp);
\draw[thick] (am)--(vp);
\draw[thick] (am)--(bp);

\node at (0,2.9) {$V^-$};
\node at (2.5,2.9) {$V^+$};
\node at (1.25,-0.6) {Bipartite graph};

\draw[-{Stealth[length=3mm]},draw=black!70,line width=1pt]
    (3.5,1.1)--node[above=5pt] {corresponds to}(5.5,1.1);

\begin{scope}[xshift=6.6cm]
\node[vertex] (u) at (0,2.2) {$u$};
\node[vertex] (v) at (2.5,2.2) {$v$};
\node[vertex] (b) at (0,0) {$b$};
\node[vertex] (a) at (2.5,0) {$a$};

\draw[query] (u)--(v);
\draw[query] (u)--(b);
\draw[query] (a)--(v);
\draw[query] (a)--(b);

\node at (1.25,-0.6) {Original graph};
\end{scope}
\end{tikzpicture}
\label{fig:bipartite-witness}
\end{figure}

Now, for intuition, suppose that this bipartite graph
$H$ is a random $d$-regular bipartite graph. Consider any two
vertices on the left side and two vertices on the
right side. These four vertices form a four-cycle
if all four edges between them are present. In a
random $d$-regular bipartite graph, the probability
of this happening is approximately $(d/n)^4$.
Since there are $\binom{n}{2}^2$ choices of such
four vertices, the expected number of four-cycles is
$
    \binom{n}{2}^2\left(\frac{d}{n}\right)^4
    =\Theta(d^4).
$
Since the graph has $nd$ edges and each
four-cycle contains four edges, the average number
of four-cycles containing an edge is
\[
    \Theta\left(\frac{d^4}{nd}\right)
    =\Theta\left(\frac{d^3}{n}\right).
\]
Thus, at least in this random setting, when $d$
is sufficiently large, there are many witnesses
and an edge belongs to many witnesses on average.

Now suppose that we find an inversion on an edge
that belongs to the average number of witnesses,
and that most of its witnesses have correctly
predicted side edges. By querying all edges
adjacent to its endpoints, we can verify the
directions of $\Theta(d^3/n)$ edges using only
$O(d)$ queries. Thus, the amortized query cost
per verified edge is
\[
    O\left(\frac{d}{d^3/n}\right)
    =O\left(\frac{n}{d^2}\right).
\]
Since there are $m=nd$ edges in our bipartite
graph, this suggests that we can verify all
edges with only $O(n^2/d)$ queries. To see why
this is useful, consider two cases. If
$d\geq n^{2/3}$, then $n^2/d\leq n^{4/3}$,
giving us the desired bound. Otherwise, if
$d<n^{2/3}$, we can simply use the algorithm
of \cite{KN}, which makes
$\tO(\sqrt{mn})=\tO(n\sqrt{d})=\tO(n^{4/3})$
queries. Thus, combining these two approaches
suggests an $\tO(n^{4/3})$ query algorithm.
Of course, this argument assumes that we can
find inversions on edges with many witnesses,
and that most of these witnesses have correctly
predicted side edges. The main technical
challenge is to guarantee these two properties,
which we address next.

\paragraph{Existence of many witnesses (non-random graphs).}
Fortunately, the existence of many witnesses is not just
a property of random graphs. \Cref{claim:bipartite-four-cycle} proves that any bipartite graph with $n$ vertices on
each side and $m$ edges contains at least
$\Omega(m^4/n^4)$ four-cycles, provided that
$m=\Omega(n^{3/2})$. (The latter condition is automatically satisfied as otherwise the $O(\sqrt{mn})$ query algorithm of \cite{KN} already takes $O(n^{1.25})$ queries.)
For $m=nd$, this gives the same
$\Omega(d^4)$ lower bound that we obtained for random
$d$-regular bipartite graphs.

However, this does not imply that every edge belongs
to many witnesses. There could be some edges that belong
to very few witnesses, while others belong to
significantly more. To deal with this, we repeatedly
remove edges that belong to fewer than
$T=\Theta(d^3/n)$ witnesses among the remaining edges.
We call these edges {\em peeled}, and the remaining
edges form the {\em core}. By definition, every edge
in the core belongs to at least $T$ witnesses.

The key observation is that we cannot peel too many
edges. Suppose that we peel $p$ edges. Every witness
consisting entirely of peeled edges can be charged
to its first peeled edge. Since each edge is peeled
when it belongs to fewer than $T$ remaining witnesses,
the total number of such witnesses is less than $Tp$.
On the other hand, our four-cycle lower bound tells
us that, if $p$ is sufficiently large, there must be
at least $\Omega(p^4/n^4)$ witnesses among the peeled edges.
Combining these bounds gives
\[
    \Omega\left(\frac{p^4}{n^4}\right) <Tp
    \quad\Longrightarrow\quad
    p = O(T^{1/3}n^{4/3}) < m/2,
\]
where the last inequality holds by choosing a suitable constant in the definition of $T$.
Thus, even in non-random graphs, at most half
of the edges are peeled, and every edge remaining
in the core belongs to $\Omega(d^3/n)$ witnesses.

\paragraph{Bounding inversions around each vertex.}
The next challenge is to ensure that most witnesses
of an inverted edge have correctly predicted side
edges. For now, we continue with the regular case,
where the degrees of all vertices in the original
graph are $O(d)$. We use a modification of an
algorithm from \cite{JWWZ} that makes
$\tO(n^2/d)$ queries and guarantees that every
vertex has at most $O(d^2/n)$ adjacent inversions. Roughly speaking, this algorithm works by going
through each vertex $v$ and sampling $\tO(n/d)$
adjacent edges looking for inversions. If an
inversion is found, we update our predictions.
As discussed earlier, this can happen only
$O(n\log n)$ times, since each inversion reduces
the solution space by a constant factor. Otherwise,
if no inversion is found, then with high probability
at most $O(d^2/n)$ of the edges adjacent to $v$
are incorrectly predicted. Some additional care is
needed to make these predictions consistent across
all vertices, but this can be done within the same
$\tO(n^2/d)$ query bound.

Now suppose that we find an inversion at an edge
$(u,v)$ in the core. Recall that this edge belongs
to at least $T=\Theta(d^3/n)$ witnesses. Since
each of $u$ and $v$ has at most $O(d^2/n)$
adjacent inversions, and each such inversion can
affect at most $O(d)$ witnesses, at most
$O(d^3/n)$ witnesses have an incorrectly
predicted side edge. By choosing the hidden constants
in our local error guarantee sufficiently small
relative to the threshold $T$, we guarantee that
at least half of the witnesses have both side
edges predicted correctly. Thus, querying all
edges adjacent to $u$ and $v$ verifies the
directions of $\Omega(d^3/n)$ edges, using only
$O(d)$ queries. This is exactly what we wanted.

\paragraph{Finding inversions in the core.}
It remains to explain how we find these inversions
efficiently. Our algorithm samples $\tO(d)$ edges
uniformly at random from the core. If we find an
inversion, we query all edges adjacent to its
endpoints, verify the opposite edges of its
witnesses, and repeat the process, including
the peeling step. Since every such expansion
verifies $\Omega(d^3/n)$ edges with $O(d)$
queries, the total query complexity of finding
and expanding inversions is
\[
    \tO\left(\frac{m}{d^3/n}\cdot d\right)
    =\tO(n^2/d).
\]
On the other hand, if we do not find an inversion
in our random sample, then with high probability
there are at most $O(m/d)=O(n)$ inversions
remaining in the core. In this case, we simply
keep the predicted directions of these edges
and defer them to be resolved later by the
algorithm of \cite{LRSZ}.

Thus, using $\tO(n^2/d)$ queries, we can verify
some edges, defer some others with at most
$O(n)$ incorrect predictions, and peel at most
half of the remaining edges. This is the main
guarantee that we need from our algorithm.

\paragraph{Handling general graphs.}
So far, we have assumed that the degrees of
all vertices are $O(d)$. In general, however,
some vertices may have much larger degrees
than others. To handle this, we partition
the vertices into buckets according to their
degrees among the currently unknown edges.
More specifically, we put vertices with degrees
between $2^{i-1}$ and $2^i$ into the same
bucket, and then partition the predicted edges
according to the buckets of their endpoints.
We process each edge bucket separately,
using a witness threshold that depends on
the maximum degrees of its two endpoints.
This allows us to apply essentially the same
argument as above, but with different degree
bounds for different buckets. By summing over
all buckets, we obtain the same guarantees,
up to logarithmic factors.

Finally, we repeatedly apply this procedure
to the peeled edges, setting $r=\Theta(n^2/m)$
when $m$ unknown edges remain. Since at most
half of the edges are peeled in each round,
we need only $O(\log n)$ rounds. We stop when
the number of remaining edges drops below
a threshold $M$, at which point we use the
prediction-based approach to handle them
with $\tO(\sqrt{nM})$ additional queries.
The total query complexity is therefore
\[
    \tO\left(\frac{n^3}{M}+\sqrt{nM}\right).
\]
Choosing $M=n^{5/3}$ balances the two terms
and gives our $\tO(n^{4/3})$ query bound.

\section{Preliminaries}\label{sec:per-vertex-bound}

For the remainder of this paper, we assume that all logs are natural (base $e$) and that $n$ and $m$ are sufficiently large to overpower constants.

In this section, we describe two algorithms from previous work that will be used to help create our algorithm. We claim no novelty for either result. The first is a modified version of an algorithm from \cite[Lemma~5.1]{JWWZ}. The original algorithm takes a graph and gives a prediction for the direction of every edge using $\tO(n^{3/2})$ queries and guarantees, with high probability, that no vertex has more than $\tO(\sqrt{n})$ adjacent errors. The modification we present uses the same techniques, but instead takes in an additional parameter $r$ and uses $O(nr \log^2n)$ queries to get a prediction with at most $d(v)/r$ errors adjacent to $v$ for every $v$ and a failure probability at most $1/n^2$. 

The following lemma gives a global prediction that incorrectly predicts few directions for each vertex. The proof more or less follows from \cite[Lemma~5.1]{JWWZ}. We present the needed modifications to their proof in \Cref{sec:apdx-local-predict}.

\begin{lemma}\label{lem:per-vertex-bound}
    Given a graph $G=(V,E)$, a set of known edge directions $Q$, a set of edges with unknown direction $E'$, and a parameter $r$. There exists a randomized algorithm $\LocalPredict(V,E,Q,E',r)$ that, with probability $1-1/n^2$ and $O(nr \log^2n)$ queries, fixes a prediction $\Epred$ of edge directions for each edge in $E'$ such that 
    \[
    \err_{\Epred}(v) \leq \frac{d(v)}{r} \hspace{1cm} \text{for every vertex } v \in V.
    \]
    Where $d(v)=\deg_{E'}(v)$ is the degree of $v$ among unknown edges in $E'$\footnote{We note that this degree among unknown edges in $E'$ is fixed at the start of the algorithm and does not change after queries reveal edge directions.}, and $\err_{\Epred}(v)$ is the number of incorrectly predicted edges in $\Epred$ adjacent to $v$. 
\end{lemma}

At a high level, the algorithm begins with a subroutine from \cite{HKK} that predicts edges and guarantees that each incorrectly predicted edge, when revealed, reduces the number of valid vertex orderings by a constant factor. This subroutine takes polynomial time and uses no queries. An algorithm that loops through all possible permutations and selects the more common edge direction would give this guarantee, but at the cost of exponential time. With recomputation, this subroutine can be wrong only $O(n\log n)$ times. Thus, we go through each vertex and sample $O(r \log n)$ adjacent edges looking for inversions. If we find one, we recompute. Otherwise, we fix a prediction for all edges adjacent to that vertex. Lastly, at the end of the prediction, we must resolve the fact that two endpoints of an edge may disagree with one another. We describe this in more detail in \Cref{sec:apdx-local-predict}.

The second algorithm we give is $\PredictionSort$ from \cite{LRSZ}. This algorithm takes in a graph and a prediction of every edge direction and determines the true direction of every edge in $O(n\log n + w)$, where $w$ is the number of incorrect predictions. We use this algorithm as a black-box and also assume the following lemma about it without proof, as it is the main result of \cite[Theorem~1.1]{LRSZ}.

\begin{lemma}\label{lem:prediction-sort}
    There is a polynomial-time randomized algorithm $\PredictionSort$ that solves the generalized sorting problem in $O(n\log n + w)$ queries with probability $1-1/n^C$ for any constant $C$.
\end{lemma}

We also note that combining $\LocalPredict$ and $\PredictionSort$ recovers the guarantees (up to $\log$ factors) of the algorithm of \cite{KN} by using the parameter $r=\Theta(\sqrt{m/n})$. If we use this parameter, we will find a prediction in $\tO(n\sqrt{m/n})=\tO(\sqrt{mn})$ queries, and the number of errors adjacent to each vertex will be at most $d(v)/r$. Summing over vertices gives at most $m/r=\tO(\sqrt{mn})$ total errors. Plugging this into $\PredictionSort$ gives a solution in $\tO(\sqrt{mn})$ queries. 
\section{Main Algorithm $\FourCycleSort$}

In this section, we give the main algorithm $\FourCycleSort$ and its helper $\Reduce$, as well as the analysis of their correctness, assuming some key lemmas that will be proven later. This section effectively serves as scaffolding for \Cref{sec:expand-witness} and \Cref{sec:many-witnesses}, which contain the main technical contributions of the paper. We begin by assuming the following lemma about a subroutine $\Reduce$ that we will prove later in \Cref{sec:many-witnesses}. 

\begin{lemma}\label{lem:reduce}
    Given a graph $G = (V,E)$, a set of known edge directions $Q$, a subset of unknown edges $E'$ given by $\FourCycleSort$ (\Cref{alg:four-cycle-sort}), and a parameter $r$. There is a randomized algorithm $\Reduce(V, E, Q, E',r)$ (\Cref{alg:reduce}) that, with probability $1-5\log^2n/n^2$ and in $O(nr\log^2n)$ queries, either peels, defers, or certifies every edge in $E'$ with the following guarantees.
    \begin{enumerate}
        \item The deferred edges come with a predicted direction that is wrong at most $n$ times.
        \item There are at most $|E'|/2$ peeled edges.
    \end{enumerate}
\end{lemma}

For $\Reduce$, we always set $r=\Theta(n^2/m)$ where $m=|E'|$ is the number of unknown edges, as already-certified edges do not affect query complexity. This gives an $O(n^3\log^2n/m)$ query algorithm that certifies some edges, defers some others with a prediction that is wrong at most $n$ times, and peels at most half of them. 

With this subroutine in hand, the structure of our overall algorithm becomes: While $|E'|$ is large, i.e., more than $M=\widetilde{\Theta}(n^{5/3})$, run $\Reduce$ on the peeled edges. This can happen at most $O(\log n)$ times. Once $|E'|$ is small, run the $\LocalPredict$ with parameter $r=\widetilde{\Theta}(\sqrt{M/n})$, which will also take $O(nr\log^2n)$ queries and add these predictions to the deferred predictions from each round of $\Reduce$. After this, run $\PredictionSort$ and return the answer. The choice of $M=\widetilde{\Theta}(n^{5/3})$ comes from a balancing between $\tO(n^3/M)$ from $\Reduce$ and $\tO(\sqrt{Mn})$ from $\LocalPredict$. In addition, the total number of incorrect predictions will be $n$ per round from $\Reduce$ and $\tO(\sqrt{Mn})$ from $\LocalPredict$, meaning $\PredictionSort$'s query complexity is balanced with that of the rest of the algorithm.

\begin{algorithm}[h]
\caption{$\FourCycleSort(V,E)$}\label{alg:four-cycle-sort}
\begin{algorithmic}[1]
    \State $E' \gets E, D \gets \emptyset$.
    \State Maintain a set $Q$ of all queried edge directions and their transitive closure.
    \State $M \gets n^{5/3}\log^{2/3}n.$
    \While{$|E'| > M$}
        \State $m \gets |E'|, r \gets 2^{25}n^2/m$.
        \State $(E',D_{\round}) \gets \Reduce(V, E, Q, E', r)$.
        \State $D \gets D \cup D_{\round}$.
    \EndWhile
    \State $\Epred \gets \LocalPredict(V, E, Q, E', \sqrt{M/n}/\log n)$.
    \State $\Epred \gets \Epred \cup D$.
    \State Resolve all edges of $\Epred$ that disagree with $Q$ and add all other verified predictions.
    \State \Return $\PredictionSort(V, E, Q, \Epred)$.
\end{algorithmic}
\end{algorithm}

Now, we examine what happens in each round of $\Reduce$ (\Cref{alg:reduce}). In \Cref{sec:techniques}, we describe the main idea behind our algorithm, which involves looking for edges that are contained in many four-cycles, then looking for inversions among such edges, and certifying many edges when these inversions are found. However, the overview given assumes a regularity condition that the degree of every vertex is at most $D$ for $D = O(m/n)$. We cannot assume this in general, but we can get the same benefits from splitting the vertices of our graph into buckets based on their degrees. Then, within each bucket, we can assume that the degrees of each vertex are upper-bounded by the threshold of that bucket. The purpose of $\Reduce$ is to give an initial prediction for the edge directions with $\LocalPredict$, then create these buckets so that they can be processed individually.

The bucketing works as follows. We first split vertices into buckets $V_0, \dots, V_{L-1}$ for $L = 2 \log n$ such that $v \in V_i$ if $B_{i-1} < d(v) \leq B_i$ where $d(v)$ is the degree of $v$ among edges in $E'$ and $B_i = 2^i$. Then, an edge $e=(u,v) \in \Epred$ is placed in bucket $\Epred_{ij}$ if $u \in V_i$ and $v \in V_j$. Also, let the size of each vertex bucket $V_i$ be $n_i$, and let the size of each edge bucket $\Epred_{ij}$ be $m_{ij}$.

Once these buckets are created, we loop through them and process each one individually with $\ProcessBucket$, then collect all peeled and deferred edges from each bucket and return them.

\begin{algorithm}[h]
\caption{$\Reduce(V, E, Q, E',r)$}\label{alg:reduce}
\begin{algorithmic}[1]
    \State $d(v) \gets \deg_{E'}(v)$ for $v \in V$.
    \State $\Epred \gets \LocalPredict(V, E, Q, E', r)$.
    \State Update $\Epred, E'$ by removing resolved edges in $Q$.

    \Statex
    \LComment{Create $L = 2 \log n$ vertex buckets.}
    \State $B_i = 2^i, V_i \gets \emptyset$ for $i \in \{0, \dots, L-1\}$.
    \State Put $v$ in $V_i$ when $B_i/2 < d(v) \leq B_i$ for $v \in V$.
    \LComment{Create $L^2$ edge buckets for $\Epred$ based on starting and ending vertex.}
    \State $\Epred_{ij} \gets \emptyset$ for $i,j \in \{0, \dots, L-1\}$.
    \State Put $e = (u,v)$ in $\Epred_{ij}$ when $u \in V_i, v \in V_j$ for $e \in \Epred$.
    \Statex
    \State $P \gets \emptyset, D \gets \emptyset$.
    \For{$i, j \in \{0, \dots, L-1\}$}
        \State $(P_{ij}, D_{ij}) \gets \ProcessBucket(V, E, Q, \Epred_{ij}, r)$.
        \State $P \gets P \cup P_{ij}, D \gets D \cup D_{ij}$.
    \EndFor
    \State \Return $(P, D)$.
\end{algorithmic}
\end{algorithm}

\section{Expanding inverted witnesses}\label{sec:expand-witness}

Before discussing the $\ProcessBucket$ algorithm, we will describe and prove an important fact about a common subroutine $\ExpandWitness$. To do this, we first define our main four-cycle gadget that we use to verify a large number of edge directions from a relatively small number of queried edges. The gadget consists of a cycle on four vertices $a, b, u, v$ where the predicted edge directions are $(u,v), (u,b), (a,v), (a,b)$. We call an instance of this gadget in our graph $G$ a \emph{witness}. Also, for a given edge $e=(u,v)$ in some subset of predicted edges $S \subseteq \Epred$, we define $w_S(e)$ as the number of witnesses containing $e$ consisting entirely of edges in $S$. In other words,
\[
    w_S(e) := \card*{\set*{(a,b) : a \neq u, b \neq v, (u,v),(a,v), (u,b), (a,b) \in S}}.
\]

Our gadget becomes useful upon finding an inversion in one of the edges. Since the gadget is symmetric, let there be an inversion between $u$ and $v$ making the true direction $(v,u)$. In this case, if we query edges $(u,b)$ and $(a,v)$ and determine that they were predicted correctly, then by transitivity we also confirm the edge $(a,b)$. See \Cref{fig:witness}.

We make the best use of this inferred edge direction in the case where there is a single edge that is contained in a large number of witnesses. If we find an inversion in this case, then by querying every edge adjacent to $u$ and $v$, we will end up inferring the \enquote{opposite edge} for each witness that had both of its \enquote{side edges} confirmed. In addition, from \Cref{lem:per-vertex-bound}, we can be sure that many of the witnesses containing $e$ have correct predictions, so all that is required is to pick a proper threshold $T_{ij}$ of witnesses containing edge $e$ where it is worth it to expand in this way. The subroutine $\ExpandWitness$ is given an edge-bucket $\Epred_{ij}$, an edge $e=(u,v) \in \Epred_{ij}'$, and the assumption both that $w_{\Epred_{ij}'}(e) \geq T_{ij}$ and that we have found an inversion at $e$ so that its true direction is $(v,u)$.

\begin{algorithm}[h]
\caption{$\ExpandWitness(V, E, Q, \Epred_{ij}', e)$}\label{alg:expand-witness}
\begin{algorithmic}[1]
   \State Let $e$ be predicted $(u,v)$ but with true direction $(v,u)$. 
   \State Query all edges in $\Epred_{ij}'$ adjacent to $u$ or $v$. 
   \State Update $Q$.
\end{algorithmic}
\end{algorithm}

We define this threshold differently for each edge bucket being processed. For each $\Epred_{ij}'$, we define $T_{ij} := 4B_iB_j/r$. We use this threshold to prove the following lemma.

\begin{lemma}\label{lem:expanding-witnesses}
    Given a graph $G=(V,E)$, a set of known edge directions $Q$, a subset of unknown edges with predicted direction $\Epred_{ij}$ given by $\ProcessBucket$ (\Cref{alg:process-bucket}), and parameter $r$. There exists a subroutine $\ExpandWitness$ (\Cref{alg:expand-witness}) that, when called on inverted edge $e = (u,v)$ with $w_{\Epred_{ij}'}(e) \geq T_{ij}$, verifies the direction of at least $T_{ij}/2$ edges in $G$ with at most $B_i + B_j$ queries. Furthermore, the total query cost of all calls to $\ExpandWitness$ for a given call to $\ProcessBucket$ is at most $O((n_i+n_j)r)$.
\end{lemma}

\begin{proof}
    Consider that we find an inversion at edge $e = (u,v)$ where $w(e) \geq T_{ij} = 4B_iB_j/r$. We then query all unknown edges in $\Epred_{ij}'$ adjacent to both $u$ and $v$, making at most $d(u) + d(v) \leq B_i+B_j$ queries. From \Cref{lem:per-vertex-bound}, we know that at most $d(u)/r$ predictions adjacent to $u$ are wrong. Each of these can affect at most $d(v)$ witnesses. Similarly, at most $d(v)/r$ predictions adjacent to $v$ are wrong, and each of these can affect at most $d(u)$ witnesses. Thus, at most $2d(u)d(v)/r \leq 2B_iB_j/r$ witnesses are affected by inversions on \enquote{side edges}. Therefore, at least $2B_iB_j/r = T_{ij}/2$ witnesses are confirmed, so we verify the direction of at least $T_{ij}/2$ edges from at most $B_i+B_j$ queries, proving the first part of the claim.

    The second part of the claim is about the total query complexity of expanding all witnesses. To do this, we first consider the query complexity of all witness expansions of edges in $\Epred_{ij}'$. Each expansion will cost us at most $B_i + B_j$ queries, but will resolve at least $2B_iB_j/r$ edges. Without loss of generality, assume that $B_i \leq B_j$; then a simple upper bound for the size of $\Epred_{ij}'$ is $|\Epred_{ij}'| \leq (n_i+n_j)B_i$. Thus, because each edge can be resolved at most once, the total queries of all expansions in $\Epred_{ij}'$ is
    \[
        (B_i+B_j)\cdot \frac{|\Epred_{ij}'|}{2B_iB_j/r} \leq (B_i+B_j) \cdot \frac{(n_i+n_j)rB_i}{2B_iB_j} \leq \frac{(n_i+n_j)r}{2}\cdot \frac{B_i+B_j}{B_j} \leq (n_i+n_j)r.\qedhere
    \]
\end{proof}
\section{Existence of many witnesses in the core}\label{sec:many-witnesses}

Now that we have shown that finding an inversion on an edge contained in many witnesses allows us to verify a large number of edges with a relatively small number of queries, we explain exactly when and how this is used during the $\ProcessBucket$ algorithm.

The algorithm is given the structure of the graph and the known information about queried edges, as well as a bucket of edges $\Epred_{ij}$ and a sampling parameter $r$. The goal is to efficiently find inversions among edges that are contained in more witnesses than the bucket's threshold $T_{ij} = 4B_iB_j/r$. To do this, we start with a process $\Peel$ which takes in the set of edges in $\Epred_{ij}$ with unknown direction, denoted $\Epred_{ij}'$, and the threshold $T_{ij}$. The process then works by peeling off edges, one by one, that are contained in fewer than $T_{ij}$ witnesses. These peeled edges get added to a set $P_{ij}$, and the \enquote{core} of edges that remain is also returned. This process takes no queries and runs in polynomial time. We will show later that, over all buckets, the number of peeled edges is small.

\begin{algorithm}[h]
\caption{$\Peel(V, E, Q, \Epred_{ij}', T_{ij})$}\label{alg:peel}
\begin{algorithmic}[1]
   \State $P_{ij} \gets \emptyset$.
   \While{there is an edge $e \in \Epred_{ij}'$ with $w_{\Epred_{ij}'}(e) <  T_{ij}$}
        \LComment{If below the threshold, remove $e$ from $\Epred_{ij}'$ and add it to $P_{ij}$.}
        \State $P_{ij} \gets P_{ij} \cup \{e\}$, $\Epred_{ij}' \gets \Epred_{ij}' \setminus \{e\}$.
   \EndWhile
   \State \Return $(\Epred_{ij}', P_{ij})$.
\end{algorithmic}
\end{algorithm}

Once edges have been peeled, finding any inversion in the core quickly will help us verify a substantial number of edges by running $\ExpandWitness$ on that edge as described in \Cref{sec:expand-witness}. Furthermore, $\ExpandWitness$ will spend $B_i+B_j$ queries for any edge in bucket $\Epred_{ij}$, so we can afford to spend $O((B_i+B_j)\log n)$ queries looking for each inversion, and the number of queries needed will only increase by an $O(\log n)$ factor. The algorithm, therefore, randomly samples $s_{ij} = 4(B_i+B_j)\log n$ edges in the core and looks for an inversion. If it finds one, it calls $\ExpandWitness$ and repeats the process, including the peeling. If it does not find one, then we ensure that the number of incorrect predictions among unresolved edges must be small, so we defer them to be resolved during $\PredictionSort$.

\begin{algorithm}[h]
\caption{$\ProcessBucket(V, E, Q, \Epred_{ij}, r)$}\label{alg:process-bucket}
\begin{algorithmic}[1]
    \State $T_{ij} \gets 4B_iB_j/r, s_{ij} \gets 4(B_i+B_j)\log n$.
    \While{there remain unresolved, undeferred, and unpeeled edges in $\Epred_{ij}$}
        \State $\Epred_{ij}' \gets $ all currently unresolved edges in $\Epred_{ij}$.
        \State $(\Epred_{ij}', P_{ij}) \gets \Peel(V, E, Q, \Epred_{ij}', T_{ij})$.
        \State Query $s_{ij}$ edges in $\Epred_{ij}'$ uniformly at random (or all of them if $|\Epred_{ij}'| \leq s_{ij}$).
        \If{inversion found at edge $e$}
            \State $\ExpandWitness(V, E, Q, \Epred_{ij}', e)$.
        \Else
            \State $D_{ij} \gets $ all unresolved edges in $\Epred_{ij}'$.
            \State \Return $(P_{ij}, D_{ij})$.
        \EndIf
    \EndWhile
    \State \Return $(P_{ij}, D_{ij})$.
\end{algorithmic}
\end{algorithm}

We now must prove three bounds about $\Peel$ and $\ProcessBucket$, namely a bound on the number of queries, a bound on the number of peeled edges, and a bound on the number of incorrect predictions among deferred edges. To start, the query complexity is easy because we already proved, in \Cref{lem:expanding-witnesses}, that the total number of queries of all calls to $\ExpandWitness$ for a given call to $\ProcessBucket$ is at most $O((n_i+n_j)r)$. Thus, since we lose an $O(\log n)$ factor querying to find these inversions in comparison to the cost of expanding them, the total query complexity of a single call to $\ProcessBucket$ is at most $O((n_i+n_j)r\log n)$. Summing over all edge buckets gives $O(nr\log^2n)$. Thus, the query complexity for a single call to $\Reduce$ is at most $O(nr\log^2n)$.

\begin{lemma}\label{lem:peel-small}
    Given a graph $G = (V,E)$, a set of known edge directions $Q$, a subset of predicted edges $\Epred_{ij}$ given by $\Reduce$ (\Cref{alg:reduce}), and a threshold $T_{ij}=4B_iB_j/r$. Let $m_{ij} = |\Epred_{ij}|$ be the fixed number of edges in the bucket before any call to $\ProcessBucket$. There is a polynomial-time algorithm $\Peel$ (\Cref{alg:peel}) that, without querying the graph, finds a core of edges $C \subseteq \Epred_{ij}'$, such that every edge $e \in C$ has $w_C(e) \geq T_{ij}$. In addition, all other edges in $\Epred_{ij}$ are either verified or peeled in $P_{ij}$ such that
    \[
        |P_{ij}| \leq \paren*{\frac{128B_iB_jn_i^2n_j^2}{r}}^{1/3} + 2(n_i\sqrt{n_j}+n_j\sqrt{n_i}).
    \]
\end{lemma}

\begin{proof}
    Examining \Cref{alg:peel}, we see that if there is any edge $e$ with $w_{\Epred_{ij}'}(e) < T_{ij}$, we peel it from $\Epred_{ij}$ and put it in $P_{ij}$. Thus, at the end of the algorithm, all edges in $\Epred_{ij}'$ have $w_{\Epred_{ij}'}(e) \geq T_{ij}$. We call the final state of $\Epred_{ij}'$ our core $C$ and need only prove that it has at least $m/2$ edges. To do this, we first prove a claim giving a lower bound on the number of four-cycles in bipartite graphs. This claim is a corollary and special case of a lemma by \cite[Lemma~1]{ErdosSimonovits1984}. We give a full proof here.

    \begin{claim}[Corollary of Lemma 1 in \cite{ErdosSimonovits1984}]
    \label{claim:bipartite-four-cycle}
        Given a bipartite graph $G \subseteq K_{a,b}$ with $a,b \geq 2$ vertices and $m \geq 2(a\sqrt{b}+b\sqrt{a})$ edges, denote the number of distinct four-cycles $C_4(G)$. It holds that
        \[
        C_4(G) \geq \frac{m^4}{32a^2b^2}.
        \]
    \end{claim}
    
    \begin{proof}
        Let the left and right sets be $L$ and $R$ with $|L|=a,|R|=b$. For each unordered pair on the left $\{u,v\} \subseteq L$, let $c(u,v)$ be the number of common neighbors between them in $R$, and let $S$ be the sum of these $c(u,v)$ for all unordered pairs $\{u,v\}$. In other words
        \[
        c(u,v) = |N(u) \cap N(v)|, \hspace{1cm} S = \sum_{\{u,v\} \subseteq L} c(u,v).
        \]
        Now, this sum $S$ counts the number of common neighbors between two distinct vertices in $L$. This is equivalent to the number of paths of length 2 from $L$ to $R$ and back to $L$. We re-evaluate this sum by counting from the right side. Each vertex $w \in R$ contributes to exactly $\deg(w) \choose 2$ such paths. Thus we also have
        \[
        S = \sum_{w \in R} {\deg(w) \choose 2}.
        \]
        Expanding and using the Cauchy-Schwarz (\Cref{prop:cauchy-schwarz}) inequality to show that $\sum_{w \in R} \deg(w)^2 \geq m^2/b$ gives
        \begin{equation}\label{eq:s-bound}
        \begin{aligned}
            S &= \sum_{w \in R} \frac{\deg(w) \cdot(\deg(w)-1)}{2} = \frac{1}{2} \sum_{w \in R} (\deg(w)^2-\deg(w)) = \frac{1}{2} \paren*{\sum_{w\in R} \deg(w)^2} - \frac{m}{2} \\
            &\geq \frac{1}{2}\cdot \frac{m^2}{b} - \frac{m}{2} \geq \frac{m^2}{4b} \geq \frac{4(a\sqrt{b}+b\sqrt{a})^2}{4b} \geq \frac{a^2b+2a^{3/2}b^{3/2}+b^2a}{b}\geq a^2. \\
        \end{aligned}
        \end{equation}    
        Now we count four-cycles. Each four-cycle is defined by two distinct paths of length two, as counted above, with common start and endpoints. Thus, each pair of vertices $\{u,v\} \subseteq L$ contributes exactly $c(u,v) \choose 2$ four-cycles. Thus, the total number of four-cycles $C_4(G)$ is
        \[
        C_4(G) = \sum_{\{u,v\} \subseteq L} {c(u,v) \choose 2}.
        \]
        Again expanding and using Cauchy-Schwarz (\Cref{prop:cauchy-schwarz}) to see that $\sum_{\{u,v\}\subseteq L} c(u,v)^2 \geq S^2/{a\choose2} \geq 2S^2/a^2$ gives
        \[
        \begin{aligned}
            C_4(G) &= \sum_{\{u,v\} \subseteq L} \frac{c(u,v) \cdot(c(u,v)-1)}{2} = \frac{1}{2} \sum_{\{u,v\} \subseteq L} (c(u,v)^2-c(u,v)) \\
            &\geq \frac{1}{2} \paren*{\frac{2S^2}{a^2} - S} \geq \frac{1}{2}\paren*{\frac{S^2}{a^2}} \geq \frac{1}{2}\cdot \frac{m^4}{16b^2}\cdot \frac{1}{a^2} \geq \frac{m^4}{32a^2b^2}.
        \end{aligned}
        \]
        Where the last line uses both the fact that $S \geq a^2$ and $S \geq m^2/4b$ from \Cref{eq:s-bound}. This proves \Cref{claim:bipartite-four-cycle}.
    \end{proof}

    With this, we continue the proof of \Cref{lem:peel-small}. We wish to show that the number of edges in the core is large, so we show that the number of edges peeled must be small. To do this, we count the number of witnesses in $P_{ij}$ and denote this as $w(P_{ij})$. We use a charging argument: for each witness in $P_{ij}$, charge it to the first edge that was peeled. For any witness, at the point that its first edge is peeled, all edges remain in $\Epred_{ij}'$. We also know that when we peel edge $e$, we have $w_{\Epred_{ij}'}(e) < T_{ij}$. These are the only witnesses that can be charged to edge $e$. Since this is true for every edge in $P_{ij}$, 
    \begin{equation}\label{eq:peel-witness-bound}
        w(P_{ij}) < T_{ij}\cdot |P_{ij}|.
    \end{equation}
    
    Now, take $P_{ij}$ and construct the following bipartite graph $G'$. For each vertex $v$, make two copies $v^-$ and $v^+$. Then, for each edge $e = (u,v)$, create an undirected edge between $u^-$ and $v^+$. This new graph $G'$ is a subgraph of $K_{n_i,n_j}$ with the same number of edges $|P_{ij}|$. Furthermore, each witness in $|P_{ij}|$ consists of edges $(a,v), (a,b), (u,v), (u,b)$, which get mapped to $(a^-,v^+),(a^-,b^+),(u^-,v^+),(u^-,b^+)$. In the bipartite graph, this is a four-cycle. Conversely, any four-cycle in $G'$ corresponds to a witness in $P_{ij}$. Thus, the number of four-cycles in $G'$ is equal to the number of witnesses in $P_{ij}$. From \Cref{claim:bipartite-four-cycle}, if $|P_{ij}| \geq 2(n_i\sqrt{n_j}+n_j\sqrt{n_i})$, then there are at least $|P_{ij}|^4/32n_i^2n_j^2$ four cycles, so
    \[
    w(P_{ij}) \geq \frac{|P_{ij}|^4}{32n_i^2n_j^2}.
    \]
    Combining this with \Cref{eq:peel-witness-bound} gives
    \begin{equation}\label{eq:dense-peel-bound}
    \begin{aligned}
        \frac{|P_{ij}|^4}{32n_i^2n_j^2} &< T_{ij}\cdot |P_{ij}|, \\
        |P_{ij}|^3 &< 32T_{ij}n_i^2n_j^2 \leq \frac{32\cdot 4B_iB_j \cdot n_i^2n_j^2}{r} \leq \frac{128B_iB_jn_i^2n_j^2}{r}.
    \end{aligned}
    \end{equation}
    Taking a cube root gives the first term in the bound in \Cref{lem:peel-small}, the second comes from the fact that the bound in \Cref{eq:dense-peel-bound} only holds when $|P_{ij}| \geq 2(n_i\sqrt{n_j}+n_j\sqrt{n_i})$. Thus, since $|P_{ij}|$ is upper-bounded in both cases, it is always upper-bounded by their sum, proving \Cref{lem:peel-small}.
\end{proof}

Next, we prove that, in total, few edges are peeled.

\begin{lemma}\label{lem:all-bucket-peel}
    Given a graph $G=(V,E)$, a set of known edge directions $Q$, a set of vertex buckets $V_i$ of size $n_i$ and edge buckets $\Epred_{ij}$ of size $m_{ij}$ for $0 \leq i, j < 2\log n$ given by $\Reduce$ (\Cref{alg:reduce}), and a parameter $r = 2^{25}n^2/m$, the total number of edges peeled across all buckets is at most $m/2$.
\end{lemma}

\begin{proof}
    This proof is just collecting terms and applying some standard combinatorial inequalities. The total number of peeled edges is equal to the sum of the sizes of all $P_{ij}$. We have
    \[
    \sum_{i,j} |P_{ij}| \leq \sum_{i,j} \paren*{\frac{128B_iB_jn_i^2n_j^2}{r}}^{1/3} + 2(n_i\sqrt{n_j}+n_j\sqrt{n_i}).
    \]
    We bound the terms individually. First, we see that
    \begin{equation}\label{eq:second-term}
    \begin{aligned}
        \sum_{i,j}2(n_i\sqrt{n_j}+n_j\sqrt{n_i}) &\leq 4\paren*{\sum_i n_i}\paren*{\sum_i\sqrt{n_i}} \leq 4n\sqrt{2\log n\cdot n} \leq M/4 \leq m/4
    \end{aligned}
    \end{equation}
    where we use both \Cref{prop:cs-consequence} to show that $\sum_i\sqrt{n_i} \leq \sqrt{2\log n \cdot n}$ and the fact that $M = n^{5/3}\log^{2/3}n$.
    Next, we bound the first term as follows.
    \begin{equation}\label{eq:first-term-rearrange}
        \sum_{i,j} \paren*{\frac{128B_iB_jn_i^2n_j^2}{r}}^{1/3} = \paren*{\frac{128}{r}}^{1/3} \cdot \sum_{i,j} (B_i^{1/3}n_i^{2/3} \cdot B_j^{1/3}n_j^{2/3}) = \paren*{\frac{128}{r}}^{1/3} \cdot \paren*{\sum_i B_i^{1/3}n_i^{2/3}}^2.
    \end{equation}
    We now bound the internal sum.
    \begin{claim}\label{claim:internal-bound}
    It holds that
        \[
            \sum_i B_i^{1/3}n_i^{2/3} \leq 4(nm)^{1/3}.
        \]
    \end{claim}
    \begin{proof}
        To prove this, we split the sum into two parts and apply H\"older's inequality (\Cref{prop:holder}) to each. We set the threshold $\Delta = 4m/n$. In the first case, we use H\"older's with $p=3$ and $q = 3/2$.
        \begin{equation}\label{eq:below-threshold}
            \sum_{B_i \leq \Delta} B_i^{1/3}n_i^{2/3} \leq \paren*{\sum_{B_i \leq \Delta} B_i}^{1/3}\cdot\paren*{\sum_{B_i\leq \Delta} n_i}^{2/3} \leq (2\Delta)^{1/3}\cdot n^{2/3} \leq 2(nm)^{1/3}.
        \end{equation}
        In the second case, where $B_i > \Delta$, we instead split the sequences differently but use the same $p=3$ and $q=3/2$ for H\"older's.
        \begin{equation}\label{eq:above-threshold}
            \sum_{B_i > \Delta} B_i^{1/3}n_i^{2/3} = \sum_{B_i > \Delta} B_i^{-1/3}(B_in_i)^{2/3} \leq \paren*{\sum_{B_i > \Delta} \frac{1}{B_i}}^{1/3}\cdot \paren*{\sum_{B_i > \Delta} B_in_i}^{2/3}.
        \end{equation}
        The first term $\sum_{B_i > \Delta} \frac{1}{B_i}$ is a geometrically decreasing sequence with ratio $1/2$, so the sum is at most $2/\Delta$. For the second term, we see that if $v \in V_i$ then $d(v) > B_i/2$ and $B_i \leq 2d(v)$. Therefore, we can rearrange our sum as
        \[
        \sum_{B_i > \Delta} B_in_i \leq \sum_i \sum_{v \in V_i} B_i \leq \sum_i \sum_{v \in V_i} 2d(v) \leq 4m.
        \]
        Plugging both of these bounds back into \Cref{eq:above-threshold} gives
        \[
        \sum_{B_i > \Delta} B_i^{1/3}n_i^{2/3} \leq \paren*{\frac{2}{\Delta}}^{1/3}\cdot (4m)^{2/3} = \paren*{\frac{2n}{4m}}^{1/3} \cdot(4m)^{2/3} = 2(nm)^{1/3}.
        \]
        Lastly, combining this with \Cref{eq:below-threshold} proves \Cref{claim:internal-bound}.
    \end{proof}
    We continue with the proof of \Cref{lem:all-bucket-peel} and substitute \Cref{claim:internal-bound} into \Cref{eq:first-term-rearrange} giving
    \[
    \sum_{i,j} \paren*{\frac{128B_iB_jn_i^2n_j^2}{r}}^{1/3} \leq \paren*{\frac{128}{r}}^{1/3} \cdot16(nm)^{2/3} = \paren*{\frac{2^{19}m^2n^2}{r}}^{1/3}.
    \]
    Lastly, substituting $r=2^{25}n^2/m$ and combining with \Cref{eq:second-term} gives
    \[
    \sum_{i,j} |P_{ij}| \leq \paren*{\frac{2^{19}m^2n^2}{2^{25}n^2/m}}^{1/3} + \frac{m}{4} \leq \paren*{\frac{m^3}{2^6}}^{1/3} + \frac{m}{4} \leq \frac{m}{2}.\qedhere
    \]
\end{proof}

Now that we have shown that the total number of peeled edges is small, it remains to show that among deferred edges, there are few incorrect predictions.

\begin{lemma}\label{lem:defer-accurate}
    Given a graph $G = (V,E)$, a set of known edge directions $Q$, a subset of predicted edges $\Epred_{ij}$ given by $\Reduce$ (\Cref{alg:reduce}), and a parameter $r$, let $m_{ij} = |\Epred_{ij}|$. At the end of $\ProcessBucket(V, E, Q, \Epred_{ij}, r)$, with probability $1-1/n^2$ the number of incorrectly predicted edge directions among deferred edges $D_{ij}$ is at most
    \[
    \err(D_{ij}) \leq \frac{m_{ij}}{B_i+B_j}.
    \]
\end{lemma}

\begin{proof}
    Assume that there are more than $m_{ij}/(B_i+B_j)$ incorrect predictions among edges in $\Epred_{ij}'$ after peeling. We sample $s_{ij} = 4(B_i+B_j)\cdot \log n$ edges in $\Epred_{ij}'$ at random, and if no inversion is found, we defer these edges. Since there are at least $m_{ij}/(B_i+B_j)$ incorrect predictions among at most $m_{ij}$ edges, the probability of finding one from a random sample is at least $1/(B_i+B_j)$. Thus, the probability of missing after $s_{ij}$ samples is at most
    \[
    \paren*{1-\frac{1}{B_i+B_j}}^{s_{ij}} \leq e^{-4\log n} \leq 1/n^4.
    \]
    If we do find an inversion, however, we expand witnesses and repeat the process. Crudely, for each call to $\ProcessBucket$ this can happen no more than $m \leq n^2$ times, as each time we verify at least one edge. Thus, by union bound, upon terminating the algorithm we must have fewer than $m_{ij}/(B_i+B_j)$ incorrect predictions with probability $1-1/n^2$, proving the claim.
\end{proof}

Lastly, we must bound the number of errors over all buckets.

\begin{lemma}\label{lem:total-deferred}
    Given a graph $G=(V,E)$, a set of known edge directions $Q$, a set of vertex buckets $V_i$ of size $n_i$ and edge buckets $\Epred_{ij}$ of size $m_{ij}$ for $0 \leq i, j < L$ given by $\Reduce$ (\Cref{alg:reduce}), and a parameter $r = 2^{25}n^2/m$, with probability $1-4\log^2n/n^2$, the number of incorrectly predicted edge directions among all deferred edges in $D$ in a given call to $\Reduce$ is at most
    \[
    \err(D) \leq n.
    \]
\end{lemma}

\begin{proof}
    From \Cref{lem:defer-accurate}, we know that the number of incorrect edges in each bucket $D_{ij}$ is at most $m_{ij}/(B_i+B_j)$ with probability $1-1/n^2$ each. By union bound, this is true of all buckets with probability at least $1-4\log^2n/n^2$. Conditioning on this and summing over all buckets gives
    \[
    \sum_{i,j} \err(D_{ij}) \leq \sum_{i,j} \frac{m_{ij}}{B_i+B_j} \leq \sum_{i,j} \frac{m_{ij}}{B_i} \leq \sum_i \sum_{v \in V_i} \frac{d(v)}{B_i} \leq n.
    \]
    The third inequality charges each edge in $m_{ij}$ to its endpoint in $V_i$, and the last inequality uses the fact that for $v \in V_i$, $d(v) \leq B_i$.
\end{proof}

We are now ready to prove \Cref{lem:reduce}.

\begin{proof}[Proof of \Cref{lem:reduce}]
    The proof follows from the combination of \Cref{lem:all-bucket-peel}, \Cref{lem:total-deferred}, the proved query complexity of $O(nr\log^2n)$, and the failure probability of $4\log^2n/n^2+1/n^2 \leq 5\log^2n/n^2$. These, together, show the guarantees we need for $\Reduce$ in \Cref{lem:reduce}.\footnote{The failure probability comes from the $4\log^2n/n^2$ in \Cref{lem:total-deferred} and the $1/n^2$ from the single call to $\LocalPredict$.}
\end{proof}
\section{Final algorithm and query bound}

\begin{lemma}[Full version of \Cref{thm:main}]
    Given an n-vertex undirected graph $G=(V, E)$ with hidden edge directions forming a DAG with a Hamiltonian path, there exists a randomized algorithm $\FourCycleSort(V,E)$ that determines, in polynomial time and with probability $1-1/n$, the direction of every edge with $O(n^{4/3}\log^{4/3}n)$ edge queries.
\end{lemma}

\begin{proof}
    First, by \Cref{lem:all-bucket-peel}, each call to $\Reduce$ peels fewer than half of the edges it receives. Since these peeled edges $P$ become the next edges passed in as $E'$, $\Reduce$ is called no more than $\log m \leq 2 \log n$ times. Furthermore, each call to $\Reduce$ costs us $O(nr\log^2 n)$ queries, and, from \Cref{lem:total-deferred}, defers a set of edges $D_{\round}$ with no more than $n$ incorrect predictions and with probability $1-4\log^2n/n^2$. Lastly, we have an additional failure probability of $1/n^2$ from the call to $\LocalPredict$.

    For each call to $\Reduce$, we set $r=\Theta(n^2/|E'|)$, since each time $|E'|$ decreases by a factor of at least two (geometrically) and never goes below $M$, the sum of these $r$ over all calls to $\Reduce$ is at most $O(n^2/M)$. Therefore, the total number of queries over all calls to $\Reduce$ is $O(n\log^2n \cdot n^2/M) = O(n^3 \log^2n /M)$. In addition, totaling up incorrect predictions gives $2n\log n$ total errors and a probability of at least $1-10\log^3n/n^2$.

    Next, by \Cref{lem:per-vertex-bound}, the call to $\LocalPredict$ on an edge set $E'$ with $|E'| \leq M$ and parameter $r = \sqrt{M/n}/\log n$ returns a set of predicted edges $\Epred$ using $O(nr\log^2n)=O(\sqrt{Mn} \log n)$ queries and failure probability $1/n^2$. This set $\Epred$ has the guarantee that, for each vertex $v$, there are at most $d(v)/r$ incorrectly predicted edges adjacent to $v$. Summing over vertices, this gives at most $M/r = O(\sqrt{Mn} \log n)$ incorrectly predicted edges in total. 

    Lastly, combining the predictions from $D$ and $\Epred$ and running $\PredictionSort$ discovers the direction of every edge in the graph in $O(\sqrt{Mn}\log n + n^3\log^2n/M)$ total queries with failure probability $1/n^C$. Setting $M=n^{5/3}\log^{2/3}n$ as in the algorithm balances this tradeoff and gives our algorithm a query complexity of $O(n^{4/3}\log^{4/3}n)$ and failure probability $1/n^2+10\log^3n/n^2+1/n^C \leq 1/n$ for $C > 1$, proving \Cref{thm:main}.
\end{proof}

\section*{Acknowledgments}

The first author thanks Moses Charikar for first introducing him to the generalized sorting problem and for many insightful conversations. We also thank Amir Azarmehr, Alma Ghafari, and Mohammad Saneian for helpful discussions over the years.

The overall approach of this paper was developed by the authors, independent of AI. However GPT-6 Astra was used in formalizing and optimizing the final bound. With no human guidance, GPT-6 Astra was not able to break the $\widetilde{O}(n\sqrt{n})$ bound of \cite{HKK} after working for 50+ hours. The authors wrote the entire paper and assume full responsibility for its content.

\printbibliography

\appendix
\section{Useful Inequalities}

\begin{proposition}[Cauchy-Schwarz]\label{prop:cauchy-schwarz}
    For $n \geq 1$ real numbers $a_1, \dots, a_n$ with sum $s = \sum_{i=1}^n a_i$, it holds that
    \[
    \sum_{i=1}^n a_i^2 \geq \frac{s^2}{n}.
    \]
\end{proposition}

\begin{proposition}\label{prop:cs-consequence}
    For $n \geq 1$ nonnegative real numbers $a_1, \dots, a_n$ with sum $s = \sum_{i=1}^n a_i$, it holds that
    \[
    \sum_{i=1}^n\sqrt{a_i} \leq \sqrt{ns}.
    \]
\end{proposition}

\begin{proof}
    To prove this, we use the Cauchy-Schwarz inequality (\Cref{prop:cauchy-schwarz}) on the sequence $\sqrt{a_1}, \sqrt{a_2}, \dots, \sqrt{a_n}$ to see that
    \[
    \begin{aligned}
        \sum_{i=1}^n a_i &\geq \frac{\paren*{\sum_{i=1}^n \sqrt{a_i}}^2}{n}, \\
        \sqrt{ns} &\geq \sum_{i=1}^n \sqrt{a_i}. 
    \end{aligned}
    \qedhere
    \]
\end{proof}

\begin{proposition}[H\"older's Inequality]\label{prop:holder}
    For positive sequences $a_k$, $b_k$ and exponents $p, q > 1$ with $1/p+1/q=1$, it holds that
    \[
    \sum a_kb_k \leq \paren*{\sum a_k^p}^{1/p}\paren*{\sum b_k^q}^{1/q}.
    \]
\end{proposition}

\begin{proposition}\label{prop:e-bound}
    For positive $a, b, c$ with $c > a$, it holds that
    \[
    \paren*{1-\frac{a}{c}}^b \leq e^{-ab/c}.
    \]
\end{proposition}

\begin{proof}
    This comes from the standard inequality $1-x \leq e^{-x}$ and simply plugging in $x=a/c$.
\end{proof}

\section{$\LocalPredict$ algorithm and analysis}\label{sec:apdx-local-predict}

In this section we give a full description of the $\LocalPredict$ algorithm and a proof of \Cref{lem:per-vertex-bound}. We again emphasize that this is only a small modification to an existing algorithm from \cite[Lemma~5.1]{JWWZ} and claim no novelty.

\begin{proof}[Proof of \Cref{lem:per-vertex-bound}]
    To prove this, we first use a claim from \cite[Lemma~3.2]{HKK} as a black box. The claim gives a polynomial-time algorithm that makes no queries and predicts the direction of every edge in the graph. The claim guarantees that if an ordering of the vertices that does not contradict the already-known information about queried edges $Q$ is selected uniformly at random, then for each edge, the probability that edge prediction agrees with the selected ordering is at least a constant. If we do not care about running time, a simple algorithm that does this is one that, for each edge $e=(u,v)$, counts the number of permutations that agree with $Q$ where $u$ comes before $v$ and the number where $v$ comes before $u$ and predicts the edge direction that occurs more often. 

    The advantage of an algorithm of this type is that it can be wrong at most $O(n\log n)$ times, adaptively. Each time we query an edge and the true direction does not match the prediction, the number of possible permutations that do not contradict $Q$ goes down by a constant factor. This happens at most $O(\log (n!))=O(n \log n)$ times, assuming we recompute the predictions each time. We give the claim implied by the polynomial-time algorithm of \cite{HKK} and take it as a black box.

    
    \begin{claim}\label{claim:poly-linear-predict-bb}
        Given an undirected graph $G=(V,E)$ with a hidden underlying DAG, and a set of known edge directions $Q$, let $\Pi_Q$ be the set of permutations of $v_1, \dots, v_n$ that do not contradict $Q$. i.e. for all $\pi \in \Pi_Q$ and $(u,v) \in Q$, we have $\pi(u) < \pi(v)$. Then, let $\Pr(u < v)$ be the probability that $\pi(u) < \pi(v)$ if $\pi$ is selected from $\Pi_Q$ uniformly at random. There is a randomized polynomial-time algorithm that makes no queries to the graph and, with probability $1-1/n^C$ for any constant $C$\footnote{The original algorithm gives a probability of $1-1/n^2$, but we can boost this to $1-1/n^C$ for any constant $C$ by simply re-running the algorithm $O(1)$ times and voting.}, assigns predicted directions to every edge in $E$ with the property that for every edge predicted in direction from $u$ to $v$, we have 
        \[
        \Pr(u < v) \geq \frac{1}{e^3}.
        \]
    \end{claim}

    From this claim, we can assume the existence of an algorithm $\Predict(V, E, Q)$ that, with probability $1-1/n^C$, returns a set of predicted edge directions $\Ehkk$ with the property that if edge $e=(u,v) \in \Ehkk$, then $\Pr(u < v) \geq 1/e^3$. Once we have this, each time we find an inverted edge $e = (u,v) \in \Ehkk$ with true direction $(v,u)$, once we add that to $Q$ we have reduced the size of $\Pi_Q$ by a constant factor of $(1-1/e^3)$. Thus, since $|\Pi_Q| \leq n!$, we know that this can happen at most $O(n \log n)$ times. For implementation purposes, we also add a parameter $E'$ such that $\Predict(V, E, Q, E')$ does not give a prediction to an uncertified edge $e$ if $e \notin E'$. This extra parameter allows us to defer and ignore sets of unknown edges, which will become important later.

    We now present the algorithm $\LocalPredict$. The algorithm works by looping through each vertex $v$ and running $\Predict(V, E, Q, E')$ on the graph, then randomly sampling $\tO(r)$ uncertified edges adjacent to $v$ uniformly at random. If an inversion is found, we repeat. If no inversion is found, then we fix the prediction for all edges adjacent to $v$ in a variable $\Epred_v$. After doing this for all vertices $v$, we may have some edges whose predictions contradict one another. What we then do is attempt to make a global prediction that sufficiently agrees with every individual $\Epred_v$. We make this global prediction the same way as before, by running $\Predict(V, E, Q, E')$. Then, we look for a vertex $v$ that disagrees with our global prediction by too much. In this case, since we know that $\Epred_v$ is accurate, most of these errors must be in our global prediction, so we can simply sample them at random and we will find an inversion quickly with high probability. We can then repeat this until our global prediction agrees sufficiently with each $\Epred_v$. After this, $\Epred$ can have error per vertex at most equal to the sum of the error of $\Epred_v$ and the disagreement between $\Epred$ and $\Epred_v$.

    We begin by analyzing the query complexity. First, we see that for each vertex, we use $O(r\log n)$ queries per round to find $\Epred_v$. Split these queries into cases where we find an inversion and cases where we don't. If we find an inversion, then we can do so at most $O(n \log n)$ times for a total of $O(nr\log^2 n)$ queries. If we do not find an inversion, we fix the prediction $\Epred_v$, and this can happen at most once per vertex for a total of $O(nr\log n)$ queries. 

    For the query complexity of the while loop, we only make $O(\log n)$ queries per round, and each time, we either find an inversion or fail. In the case that we don't fail, this can again happen at most $O(n\log n)$ times for a total of $O(n \log^2 n)$. Overall, we use $O(nr\log^2 n)$ queries.

    \begin{algorithm}[H]
    \caption{$\LocalPredict(V, E, Q, E', r)$}\label{alg:local-predict}
    \begin{algorithmic}[1]
        \State $\Epred_v \gets \emptyset$ for all $v \in V$.
        \For{$v \in V$}
            \While{$\Epred_v$ not set or inversion found}
                \State $\Ehkk \gets \Predict(V, E, Q, E')$.
                \State $\Epred_v \gets $ all edges in $\Ehkk$ adjacent to $v$.
                \State Randomly sample $12r\log n$ uncertified edges of $E'$ adjacent to $v$ and query them.
                \State Update $Q$.
            \EndWhile
        \EndFor
        \While{There is some $v$ where $\Ehkk$ disagrees with $\Epred_v$ at least $2d(v)/3r$ times}
            \State $\Ehkk \gets \Predict(V,E,Q, E')$.
            \LComment{Let $d(v)$ be the degree of $v$ among unverified edges in $E'$, fixed at the start of the algorithm.}
            \If{$\Ehkk$ disagrees with $\Epred_v$ at least $2d(v)/3r$ times for some $v$}
                \State Randomly sample $8 \log n$ such edges and query them.
                \State Update $Q$.
                \If{no inversion found}
                    \State \Return $\mathsf{FAIL}$.
                \EndIf
            \EndIf
        \EndWhile
        \State \Return $\Ehkk$.
    \end{algorithmic}
    \end{algorithm}

    Next we prove the claim that, with high probability, every $\Epred_v$ is accurate.

    \begin{claim}\label{claim:per-vertex-ev}
        Given a graph $G = (V,E)$, a set of known edge directions $Q$, a set of unknown edge directions $E'$, and a parameter $r$. After running $\LocalPredict(V, E, Q, E', r)$, with probability $1-1/3n^2$, for each vertex $v \in V$ we have
        \[
        \err_{\Epred_v} \leq \frac{d(v)}{3r}
        \]
        where $\err_{\Epred_v}$ is the number of incorrectly predicted edges in $\Epred_v$.
    \end{claim}
    \begin{proof}
        We start by examining a single prediction attempt on a single vertex. We sample $12r \log n$ edges uniformly at random, and if there are more than $d(v)/3r$ incorrectly predicted edges out of a maximum of $d(v)$ uncertified edges, then each sample finds an inversion with probability at least $1/3r$, so the probability of finding none is at most 
        \[
        (1-1/3r)^{12r\log n} \leq e^{-4 \log n} \leq 1/n^4,
        \]
        where we make use of the basic fact of \Cref{prop:e-bound}. In total, this subroutine can happen at most $O(n\log n)$ times, so by union bound, the probability that we always find an inversion when there are at least $d(v)/3r$ is at least $1-1/3n^2$. Thus, for each $v$, if we condition on this high-probability event, we can be sure that each $\Epred_v$ has at most $d(v)/3r$ errors, proving \Cref{claim:per-vertex-ev}. 
    \end{proof}

    Now that we have accurate $\Epred_v$ for each $v$, we must verify that the global prediction works as intended. First, in the case where $\Ehkk$ disagrees with $\Epred_v$ by at most $2d(v)/3r$ for every $v$, we simply return $\Ehkk$, and each vertex has error at most $2d(v)/3r+d(v)/3r = d(v)/r$. Assume there is some $v$ where this is not the case, so we have at least $2d(v)/3r$ disagreements. Each such disagreement between $\Ehkk$ and $\Epred_v$ causes at least one of them to be wrong, but we know that $\Epred_v$ is wrong at most $d(v)/3r$ times. Thus, $\Ehkk$ is wrong in the majority of these disagreements. If we sample $8 \log n$ such disagreements, then the probability that we find none where $\Ehkk$ is wrong is at most 
    \[
    (1-1/2)^{8\log n} \leq e^{-4 \log n} \leq 1/n^4.
    \]
    Thus, by a union bound over $O(n\log n)$ possible predictions for $\Ehkk$, we fail with probability at most $1/3n^2$.

    Combining this with \Cref{claim:per-vertex-ev}, we see that $\LocalPredict$ finds a prediction $\Epred = \Ehkk$ for which each vertex has error at most $d(v)/r$, and uses $O(nr\log^2 n)$ queries. Furthermore, the probability of failure is $1/3n^2+1/3n^2$ plus an additional $n^{-C}$ for each call to $\Predict(V, E, Q, E')$. Simply taking $C \geq 4$ gives a failure probability at most $1/3n^2$ over a union bound to all calls to $\Predict$. This gives a total failure probability of $1/3n^2+1/3n^2+1/3n^2 =1/n^2$, proving \Cref{lem:per-vertex-bound}.
\end{proof}

\end{document}